\documentclass{elsarticle}

\usepackage{graphicx}
\usepackage{amssymb}
\usepackage{amsmath}

\begin{document}

\title{Clustering of tag-induced sub-graphs in complex networks}

\author[k]{P{\'e}ter Pollner}
\author[k]{Gergely Palla}
\author[k,e]{Tam\'as Vicsek}

\address[k]{Statistical and Biological Physics Research Group of HAS, E{\"o}tv{\"o}s University, 1117 P\'azm\'any P. Stny 1/A, Budapest, Hungary}
\address[e]{Department of Biological Physics, E{\"o}tv{\"o}s University, 1117 P\'azm\'any P. Stny 1/A, Budapest, Hungary}

\begin{abstract}
We study the behavior of the clustering coefficient in tagged networks.
The rich variety of tags associated with the nodes in the studied systems 
provide additional information about the 
entities represented by the nodes which can be important for
practical applications like searching in the networks. 
Here we examine how the clustering coefficient changes when
narrowing the network to a sub-graph marked by a given tag, and
how does it correlate with various other properties of the sub-graph.
Another interesting question addressed in the paper is how the
clustering coefficient of the individual nodes is affected by the
tags on the node. We believe these sort of analysis help acquiring
 a more complete description of the structure of large complex systems.
\end{abstract}

\begin{keyword}
networks \sep tags \sep clustering
\PACS 02.70.Rr 
\sep
89.20.-a 
\sep
89.75.Hc 
\end{keyword}

\maketitle

\section{Introduction}
A wide range of complex natural, social and technological
phenomena can be analyzed in terms of \emph{networks} capturing 
the intricate web of connections among the units (building blocks) of 
the system under study \cite{Laci_revmod,Dorog_book}. Over the last
decade it has turned out that  networks corresponding to realistic systems
can be highly non-trivial, characterized by  a low average distance
combined with a high average clustering coefficient \cite{Watts-Strogatz},
anomalous degree distributions \cite{Faloutsos,Laci_science} and an intricate
modular structure \cite{GN-pnas,CPM_nature,Fortunato_report}. Although 
the majority of complex network studies concern simply the topology of the 
graph corresponding to the investigated system, there is a steadily increasing 
interest towards \emph{tagged networks} as well. 

The inclusion of \emph{node tags}
(also called as attributes, annotations, properties, categories, features)
 leads
 to a richer structure, opening up the possibility for a more comprehensive
 analysis.
These tags can correspond to basically
 any information about the nodes and in most cases a single node
 can have several tags at the same time.
 The appearance of tags
e.g., in biological networks is very common 
\cite{Mason_nets_in_bio,Zhu_nets_in_bio,Aittokallio_nets_in_bio,Finocchiaro_cancer,Jonsson_Bioinformatics,Jonsson_BMC}, 
where they usually refer to the biological function of the units represented
 by the nodes (proteins, genes, etc.).
 Another field of high interest and
special importance from the point of view of practical applications 
is given by folksonomies and collaborative tagging systems
 like CiteUlike, Delicious or Flickr \cite{Cattuto_PNAS,Lambiotte_ct}. 
These originate from users associating tags to certain objects 
(web-pages, photos, etc.), with each tagging action defining a user-tag-object
 triplet. The natural representation of these systems is given by
 tri-partite graphs, or in a more general framework by \emph{hypergraphs}
 where the hyperedges can connect more than two nodes together. Modeling 
folksonomies with random hypergraphs is a very interesting new field in
complex network theory which is likely to gain serious importance in the 
close future \cite{Newman_PRE,Caldarelli_PRE}.
Interesting applications of node features can be seen in the studies
of \emph{co-evolving} network models as well, where the evolution of 
the network
topology affects the node properties and vice versa
\cite{Zimmermann_coevlov,Eguiluz_coevolv,Watts_science,Ehrhardt_coevolv,Newman_coevolv,Gil_coevolv,Vazquez_PRE,Vazquez_cond_mat,Kozma_coevolv,Benczik_coevolv}.
These models are aimed at describing
 the dynamics of social networks, in which people with similar opinion
 are assumed to form ties more easily, and the opinion of connected people
 becomes more similar in time. 

Finally, we mention our previous work 
aimed at the fundamental statistical features of tagged networks where
the tags are organized into an ontology \cite{our_tagged}. 
According to our results
an interesting self-similarity and scaling can be observed in the link density
of the sub-graphs spanning between nodes marked by a given tag
or any descendants of this tag in the ontology. Here we continue the study
of the relationship between the distribution of tags and the topology by
focusing on the \emph{clustering coefficient} in tagged networks.
 The clustering coefficient, $C$ is an important measure of the
transitivity in a network, measuring the probability of two neighbors of 
the same node being linked to each other as well \cite{Watts-Strogatz}. 
The average $C$ of real networks is usually significantly higher than
that of a corresponding Erd\H{o}s-R{\'e}nyi (E-R) random graph \cite{E-R}, 
and for some networks $C$ was
claimed to scale as $d^{-1}$ with the node degree $d$ \cite{Laci_hier_scale}. 

Related to that, here we shall investigate how the number of tags on 
the node affect the clustering  coefficient. Furthermore, we shall also
study the clustering coefficient in the tag-induced sub-graphs. These 
sub-graphs can be important when e.g., searching in the network. The
narrowing or widening of the specificity of the tag according to which we
are searching corresponds to switching between sub-graphs embedded in
 one an other, which is presumably accompanied by the change in the 
cohesiveness of the sub-graph in question. We shall study this change
in the cohesiveness via the clustering coefficient. Tagged networks and
the change in the cohesiveness of a chosen sub-graph due to the 
narrowing/widening of the tag specifying the included nodes can be
important in the usage of \emph{recommendation systems} as well. 
These systems are aimed at offering new services to customers based on
previously purchased services. A natural representation of the 
available services is given by a network with link weights corresponding to 
the frequency of simultaneous purchasing of the two items, and above
a certain number of services the service providers usually organize
the services into a hierarchy which can be represented by node tags. 
The widening or narrowing of the scope of services to take into
account during the recommendation process is similar to switching
between more specific or more general tags when searching in a
tagged network.

 The paper is organized
as follows: in Sect.\ref{sect:defs}. we specify the two alternative
definitions mostly used for the clustering coefficient as well as
the various quantities related to the tag-induced sub-graphs. 
In Sect.\ref{sect:apps}. we briefly describe the studied networks and 
show the results concerning the behavior
of the clustering coefficient, and finally we 
conclude in Sect.\ref{sect:concl}.

\section{Definitions}
\label{sect:defs}

\subsection{Clustering coefficient}
The clustering coefficient has actually two (slightly different) definitions, 
one was given by D.J. Watts and S.H. Strogatz based on the 
local neighborhood around a given node \cite{Watts-Strogatz}.
In this approach, the clustering coefficient, $C_i$ 
of node $i$ is given by
\begin{equation}
C_i\equiv\frac{2t_i}{d_i(d_i-1)},
\label{eq:clust_def}
\end{equation}
where $t_i$ denotes the number of triangles passing through $i$ (equivalent
 to the number of links between the neighbors of $i$), and $d_i$ is the 
degree of node $i$. (The clustering coefficient of nodes with less then 
two links is zero by definition). The clustering coefficient of a sub-graph
 (or the whole network) is simply $\left< C\right>$  averaged over
its nodes. 

To avoid ambiguity, we shall refer to as the \emph{transitivity coefficient} for
the alternative definition of the clustering coefficient, given only
 for sub-graphs and not for the individual nodes. The transitivity
 coefficient $T$ of a sub-graph $G$ is given by
\begin{equation}
T=\frac{3 t_{G}}{b_G},
\end{equation}
where $t_G$ denotes the number of triangles in the sub-graph and $b_G$ 
stands for the number of connected triples of nodes (equivalent to the number
of paths with length two) \cite{Barrat_clust,Newman_WS}. 
The factor of 3 in the numerator accounts for the 
fact that each triangle contributes to 3 connected triples of nodes, 
one for each of its 3 nodes. The main difference between the two definitions
is that (\ref{eq:clust_def}) tends to weight the contribution from low degree
nodes more heavily, because such nodes have a smaller denominator 
\cite{Newman_SIAM}.

\subsection{Tag frequencies}
As we mentioned in the Introduction, the number of associated tags 
can vary from one node to the other, and similarly, the 
\emph{frequency} of the different tags can also be
 rather heterogeneous. What can make the picture  more complex is that 
in many systems the tags refer to \emph{categories}
of a \emph{taxonomy} or \emph{ontology} (capturing the view of a 
certain domain, e.g., protein functions). This means that the tags are
organized into a structure of relationships
which can be represented by a directed acyclic graph (DAG), in which a 
directed link from a category $\alpha$ pointing to another category 
 $\beta$ represents a ``$\beta$ is a sub-category of $\alpha$'' relation. 
(Note that a given sub-category can have more than one in-neighbors in 
 the DAG.) The nodes close to the root in the DAG are usually related to
 general properties, and as we follow the links towards the leafs, the
 categories become more and more specific.

Given the DAG between the possible tags, we can define
the frequency of a given tag $\alpha$ in two
different ways \cite{our_tagged}:
\begin{eqnarray}
p_{\alpha}&\equiv& N_{\alpha}/N, \\
\tilde{p}_{\alpha}&\equiv& \tilde{N}_{\alpha}/N,
\label{eq:tag_freq}
\end{eqnarray}
where $N_{\alpha}$ denotes the number of nodes tagged with $\alpha$,
$\tilde{N}_{\alpha}$ stands for the number of nodes tagged with $\alpha$ or
any of its descendants, and $N$ is equal to the total number of nodes in
 the network. 
Low frequency tags are more specific in an information theoretical sense,
whereas high frequency tags carry almost no information (e.g., being
 tagged by the root in the annotation DAG adds absolutely no information
 to the description of a node). The $\tilde{p}_{\alpha}$ plays an important
role in \emph{semantic similarity measures} \cite{Resnik,Lin}, e.g., in case 
of the similarity measure defined by P. Resnik the similarity of two
tags is given as $-\log\tilde{p}$ of their lowest common ancestor in the DAG.

\subsection{Tag-induced sub-graphs}
\label{sect:tag-induced}
\begin{figure}[ht]
\centerline{\includegraphics[width=\textwidth]{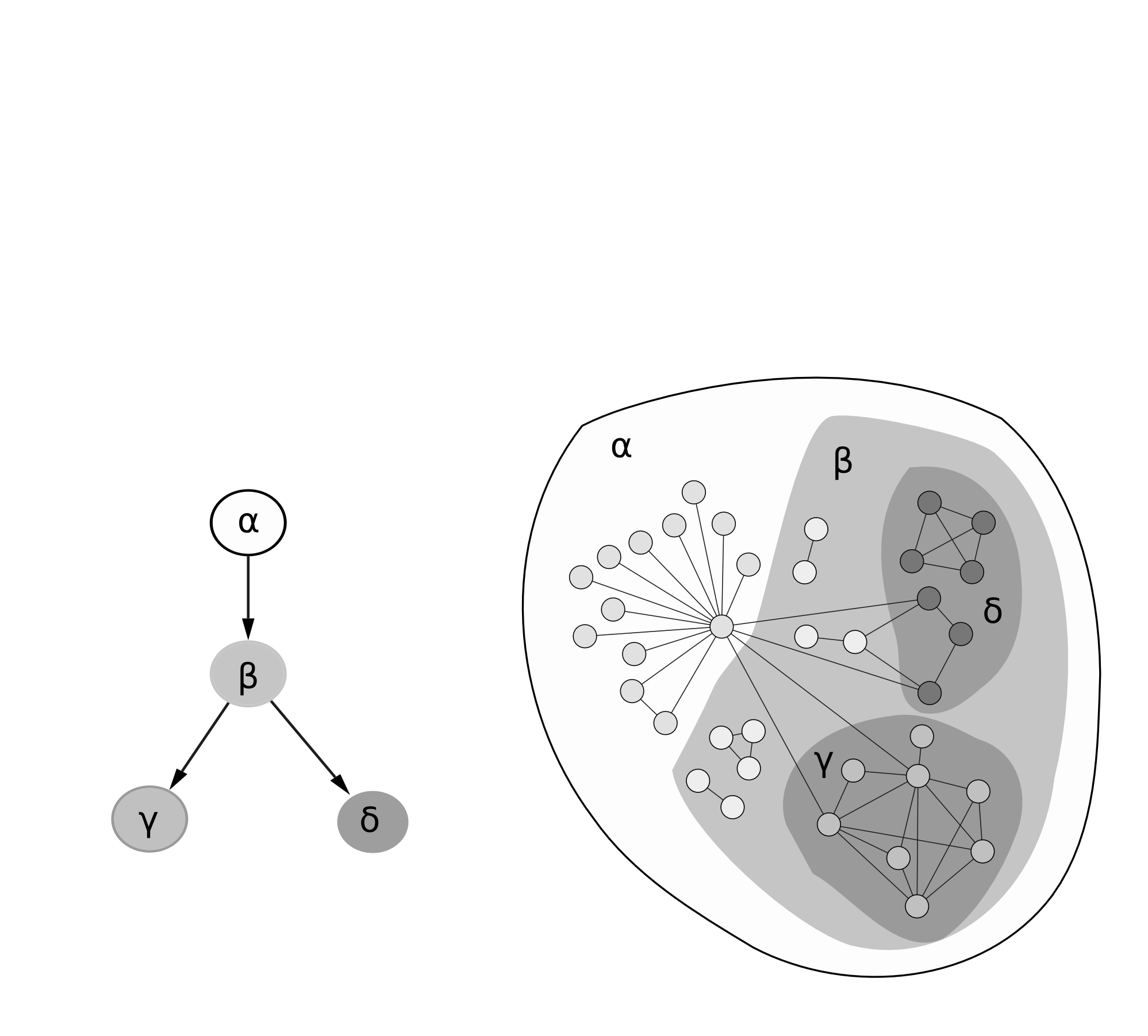}}
\caption{Illustration of the concept of tag-induced sub-graphs. On the 
left we show a part from the DAG of the tags in the MIPS network. The induced
sub-graphs are displayed on the right, following the same color coding
as the tags on the left.}
\label{fig:illustr}
\end{figure}

One of the key objects of the present study is given by the
\emph{tag-induced sub-graphs}, spanning between nodes marked
 by a given  tag $\alpha$ and any of its descendants.
(For an illustration see Fig.\ref{fig:illustr}).
 The number of nodes 
in this sub-graph is given by $\tilde{N}_{\alpha}$, whereas the number 
of links can vary between $\tilde{M}_{\alpha}=0$ and
$\tilde{M}_{\alpha}=\tilde{N}_{\alpha}(\tilde{N}_{\alpha}-1)/2$. 
According to our previous results, an interesting self-similar property
and scaling can be observed when comparing the different tag-induced
 sub-graphs \cite{our_tagged}. The expected number of links in the 
tag-induced sub-graphs follows a power-law in function of the number of nodes
in the sub-graphs, $\tilde{M}\sim \tilde{N}^{\mu}$, 
characterized by an exponent $\mu$ related to the 
\emph{tag-assortativity}. A network is tag-assortative, if nodes with similar
tags are linked with a larger probability than at random. 

In the 
absence of correlations between the tags and the graph topology 
the mentioned tag-assortativity exponent equals $\mu=2$. 
(In this case a tag-induced sub-graph corresponds to just a 
random sample from the network with $\tilde{N}(\tilde{N}-1)/2$ possible
places for links which are filled with a uniform probability 
independent of the sub-graph size $\tilde{N}$). In contrast, for
 the studied systems $\mu$ was found to be between 1 and 1.5,
which is a signature of tag-assortativty as we shall see shortly. The 
probability to find a link between a randomly chosen pair of nodes in the 
tag-induced sub-graph denoted by $\rho$ scales as $\rho\sim\tilde{M}/\tilde{N}^{2}\sim\tilde{N}^{\mu-2}$. When $\mu<2$, this linking probability $\rho$ 
becomes larger for the smaller sub-graphs, corresponding to more specific tags,
thus, the network is tag-assortative. 

Based on the above behavior one expects that 
the clustering coefficient (transitivity coefficient) in the tag-induced
 sub-graphs of more specific tags should be higher on average as well.
In an Erd\H{o}s-R{\'e}nyi (E-R) graph \cite{E-R} with the same number of nodes 
and links
as a chosen tag-induced sub-graph the clustering coefficient would
be equal to the linking probability $\rho$. Thus, we expect $\left< C\right>$
 (and $T$) to grow at least as $\tilde{N}^{\mu-2}$ on average when moving
from the tag-induced sub-graph of a general tag to the tag-induced
sub-graph of a more specific one.

\section{Applications}
\label{sect:apps}

We studied the behavior of the clustering coefficient in the 
same three networks of high interest as in \cite{our_tagged}, capturing the 
relations between interacting proteins, collaborating scientists, 
and pages of an on-line encyclopedia. The protein-protein interaction 
network of MIPS \cite{MIPS} consisted of $N=4546$ proteins, 
connected by $M=12319$ links, and the tags attached to the nodes corresponded 
 to $2067$ categories describing the biological processes the proteins
 take part in. The DAG between these categories was obtained from
 the Genome Ontology database \cite{GO}. 

The co-authorship network originated from MathSciNet 
(Mathematical review collection of the American Mathematical Society)
 \cite{Mathscinet}, with $N=391529$ scientists connected by 
$M=873775$ links of collaboration. The node tags were obtained from the $6499$
 different subject classes of the articles, which were organized
 into a DAG. Thus, the set of tags attached to each author was 
the union of all subject-classes that appeared on her/his papers.

Finally, the third network was given by a subset of pages from the English 
Wikipedia \cite{Eng_wiki,Zlatic_wikipedia,Capocci_wiki_PRE,Capocci_wikipedia},
connected by hyperlinks 
embedded in the text of the pages. At the bottom of each page, one can find
 a list of categories, which were used as node tags. Since each 
 wiki-category is a page in the Wikipedia as well, we removed these 
pages from the network to keep a clear distinction between nodes 
and attributes. Furthermore, we kept only the mutual links between 
the remaining pages. Similarly
 to the biological processes in the MIPS network or the subject
 classes in the MathSciNet, the wiki-categories can have sub-categories
 and are usually part of a larger wiki-category. However, when representing
 these relations as a directed graph, some directed loops appear, 
 therefore, they do not form a strict DAG. Thus, we removed a 
few relations from this graph until it turned into a DAG,
 following a method detailed in the Appendix of \cite{our_tagged}. 
The chosen subset of pages corresponded to the tag-induced sub-graph of
 ``Japan'', consisting of $N=43307$ nodes, $M=102753$ links with 
$3197$ sub-categories appearing as tags on the nodes.

\begin{figure}[ht]
\centerline{\includegraphics[width=8cm]{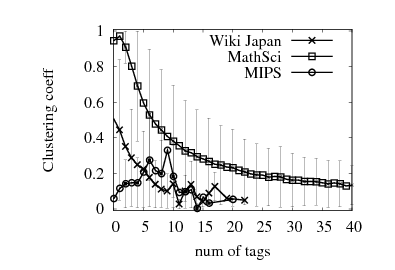}}
\caption{The average clustering coefficient of nodes in function of their
number of tags for the Wiki-Japan (crosses), the MathSciNet (boxes) and
the MIPS network (circles). }
\label{fig:num_of_tags}
\end{figure}
In Fig.\ref{fig:num_of_tags}. we show the average clustering coefficient
of the nodes, $\left< C\right>$ in function of their number of tags.
As we already pointed out, the studied systems are all tag-assortative,
thus, neighboring nodes are likely to share common tags. From this
it follows that nodes with a lot of tags are likely to have different
neighbors which are not linked to each other, thus,
we expect the clustering coefficient of such nodes to be lower than those
with few tags. We see this expected decreasing tendency with some 
fluctuations towards the large
number of tags for the Wiki-Japan network. 
In case of the MathSciNet the overall tendency is 
decreasing as well, however, the $\left< C\right>$ of nodes with one tag 
is larger than that of those with none, producing a maximum at 
the beginning of the curve. Most peculiar is the curve of the MIPS (circles),
showing an increasing tendency for low number of tags and a fluctuating
plateau for larger values. This network has shown interesting differences from
the other two networks in our previous study as well \cite{our_tagged}. 
E.g., hubs with a rather special function (described by a single- or 
only few tags) could be found, contradicting the simple argument of
large node degree correlating with large number of tags on the node.
(The proteins helping other proteins to fold are good examples for this).
 Naturally, the clustering coefficient is expected to be low for 
these nodes due to the large degree. 

\begin{figure}[t]
\centerline{\includegraphics[width=8cm]{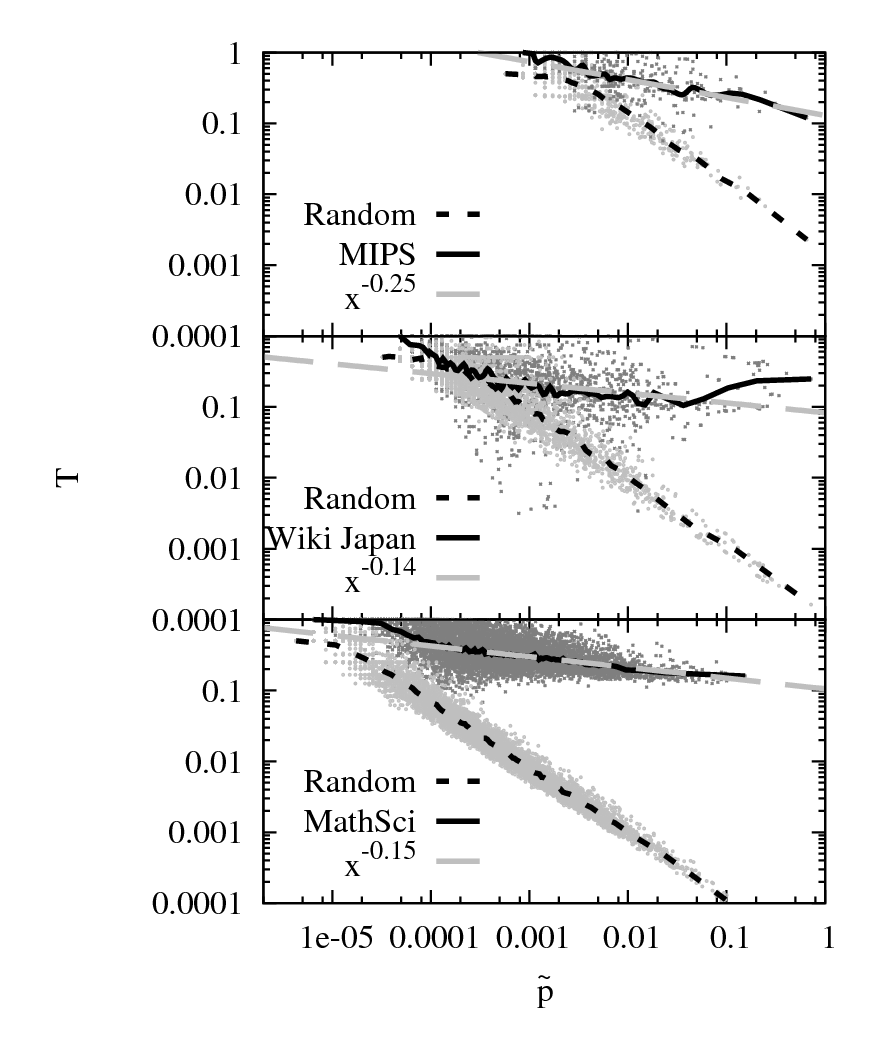}}
\caption{The transitivity coefficient of tag-induced sub-graphs in
function of their relative size, $\tilde{p}$. In each panel,
the dark symbols show the measured value, whereas the gray symbols correspond
to the estimated value in an E-R graph with the same number of nodes
and links as the tag-induced sub-graph. The continuous- and dashed black 
curves show the average values. For each network there is a sub-range of
$\tilde{p}$ values in which the average of the measured $T$ decays
more or less as a power-law, this is shown by the dashed gray lines. 
}
\label{fig:sub_clust}
\end{figure}
In Fig.\ref{fig:sub_clust}. we display the transitivity coefficient of 
the tag-induced sub-graphs in function of $\tilde{p}$ corresponding
to their relative size. For each network we 
also plotted the values one would obtain in E-R graphs with the same
number of nodes and links as the tag-induced sub-graphs, showing
the $\tilde{N}^{\mu-2}$ scaling as discussed in Sect.\ref{sect:tag-induced}.
For all three networks the average $T$ is clearly higher than what
we would get in the E-R counterpart, which is a signature of
correlations making these sub-graphs more cohesive. This difference
becomes really significant for the larger sub-graphs (more general tags).
 Although the 
average of the actual $T$ of the tag-induced sub-graphs can be fitted 
with a power-law only in a limited
range, an apparently decreasing $\left< T \right>$ curve can be observed 
in function of $\tilde{p}$ for the MIPS and the MathSciNet. 
In case of the Wiki-Japan the overall nature 
of the same curve is decreasing as well with an increasing tail at large
$\tilde{p}$ values. 
This may be an effect of the poorer statistics (smaller number
 of sub-graphs) in this region. The exponents of the fitted power-laws are
between -0.1 and -0.3, thus, the decrease in the transitivity with increasing
$\tilde{p}$ is quite slow and the larger tag-induced sub-graphs remain 
rather clustered on average.

\begin{figure}[hbt]
\centerline{\includegraphics[width=\textwidth]{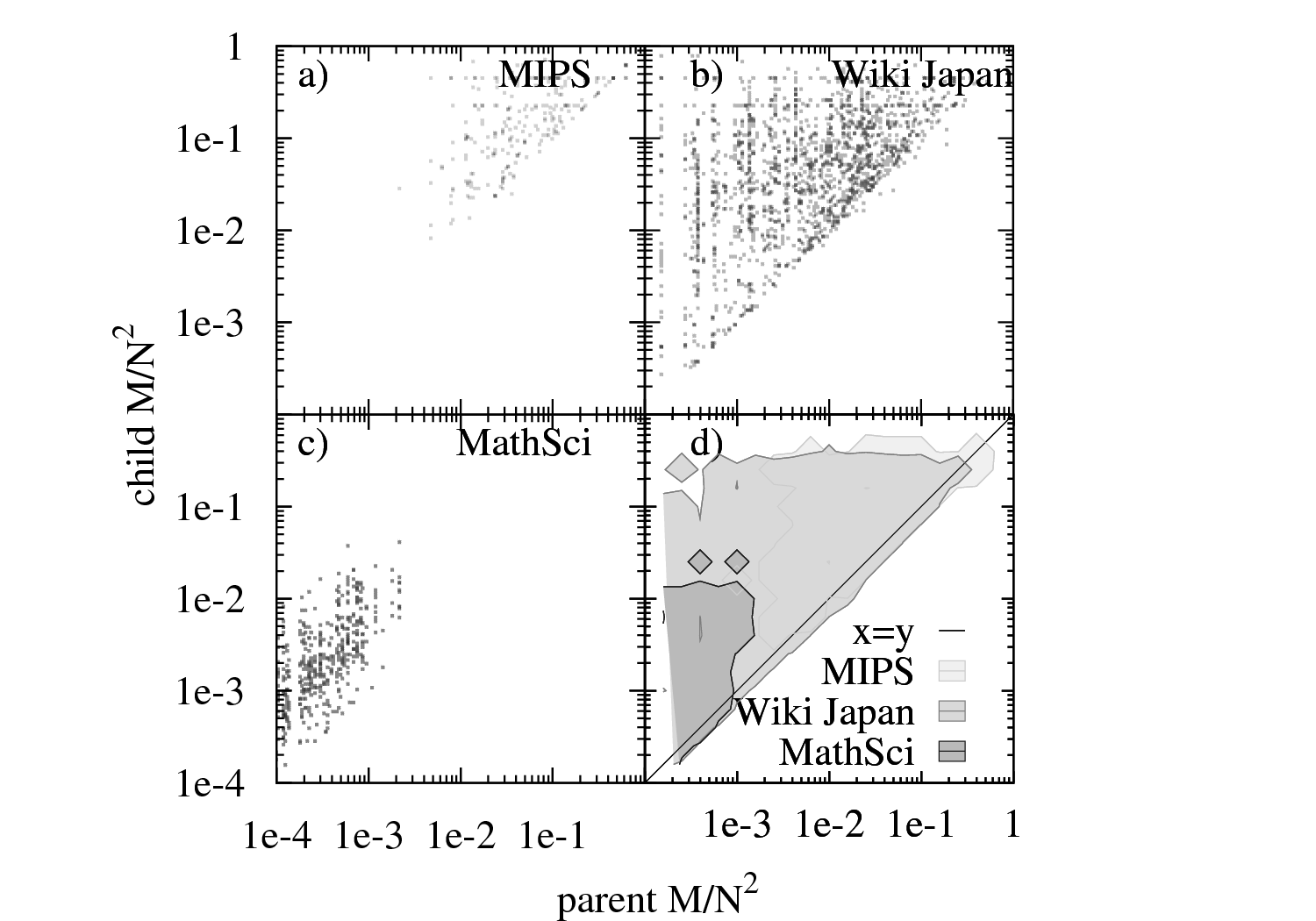}}
\caption{
Scatter plot of the linking probability $\rho\sim M/N^2$ of the tag-induced
sub-graph in function of the $\rho\sim M/N^2$  in the tag-induced sub-graph
of the direct ancestor (``parent'' in the DAG) of the tag for the MIPS (a), the
Wiki-Japan (b) and the MathScinet networks (c). (Each point corresponds
to a direct ancestor-descendant pair.) 
Panel (d) depicts a contour plot of the point 
densities obtained from the scatter plots.}
\label{fig:child_rho}
\end{figure}
Next, we aim at investigating the direct ancestor-descendant relation
of the tags by comparing various statistics of the corresponding
tag-induced sub-graphs. In other words, we study how the sub-graphs
 change if we narrow our field of interest by moving in the DAG of tags along
a directed link to a more specific tag. 
In Fig.\ref{fig:child_rho}. we simply plot
the link probability $\rho\sim M/N^2$ of the descendant 
(``child'')  in function of the $\rho$ of its direct ancestor (``parent'').
As expected, the vast majority of the points falls above the $y=x$ curve, 
corresponding to a larger linking probability in the descendants sub-graph.
Some differences can be observed between the three systems which are
emphasized in Fig.\ref{fig:child_rho}d showing a contour plot of the
point densities obtained from the scatter plots Fig.\ref{fig:child_rho}a-c:
For the Wiki-Japan the points spread all over the range above the
$y=x$ line, thus, an ancestor having a low $\rho$ value can still have
 a direct descendant with a high $\rho$. In case of the other two networks
the points remain closer to the diagonal, corresponding to more correlated
``child-parent'' $\rho$ values.

\begin{figure}[hbt]
\centerline{\includegraphics[width=\textwidth]{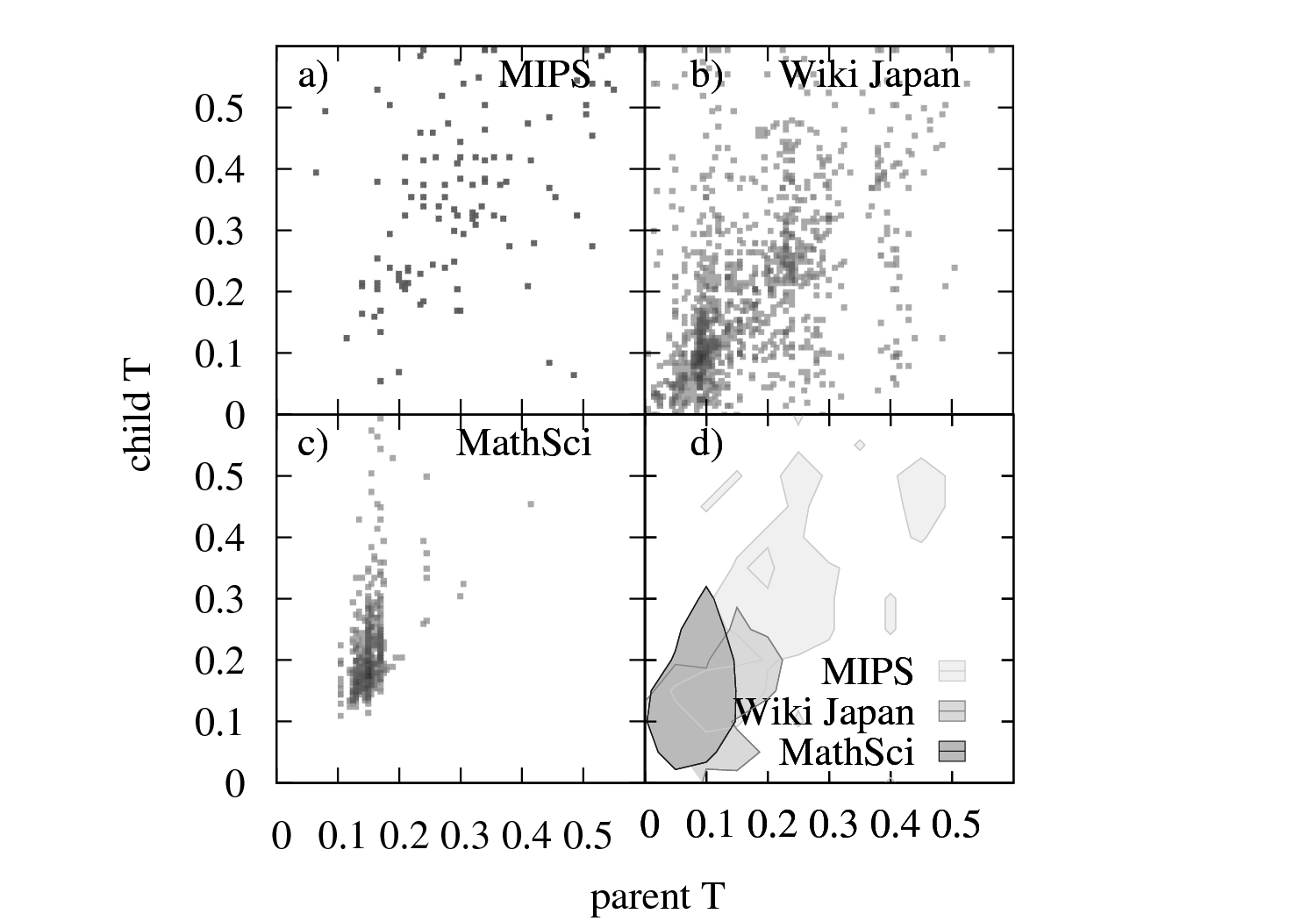}}
\caption{
 The transitivity coefficient of the tag-induced sub-graph in function 
of the transitivity coefficient of the tag-induced sub-graph of its 
direct ancestor for the MIPS (a), the Wiki-Japan (b) 
and the MathScinet networks (c). (Each point corresponds
to a direct ancestor-descendant pair.) For a comparison, in panel (d) we show 
a contour plot of the point densities obtained from the scatter plots.}
\label{fig:child_T}
\end{figure}
In Fig.\ref{fig:child_T}. we show the transitivity coefficient
 $T$ in the induced sub-graph of the descendants in function of $T$ of 
their direct ancestors induced sub-graph. The majority of the points falls
above the $y=x$ line, thus, the increase of $T$ when moving from
the ancestor to the descendant is more common than the decrease. This
 tendency is most pronounced in case of the MathSciNet. 
In case of the MIPS and especially the Wiki-Japan we can also 
find numerous ancestor-descendant pairs where $T$ is 
actually smaller for the descendant. By examining such pairs in more details
it turned out that the usual cause for this effect is another descendant
of the ancestor having a high transitivity: in such settings the transitivity
of the ancestor becomes roughly the average of the transitivities of its
descendants.
 According to the contour plot 
(Fig.\ref{fig:child_T}d) obtained from the individual scatter plots 
the transitivity coefficients of the direct ancestor-descendant pairs are 
more correlated
than e.g., the corresponding linking probabilities, as the majority of the
 points in the scatter plots gather around the diagonal.

\begin{figure}[hbt]
\centerline{\includegraphics[width=\textwidth]{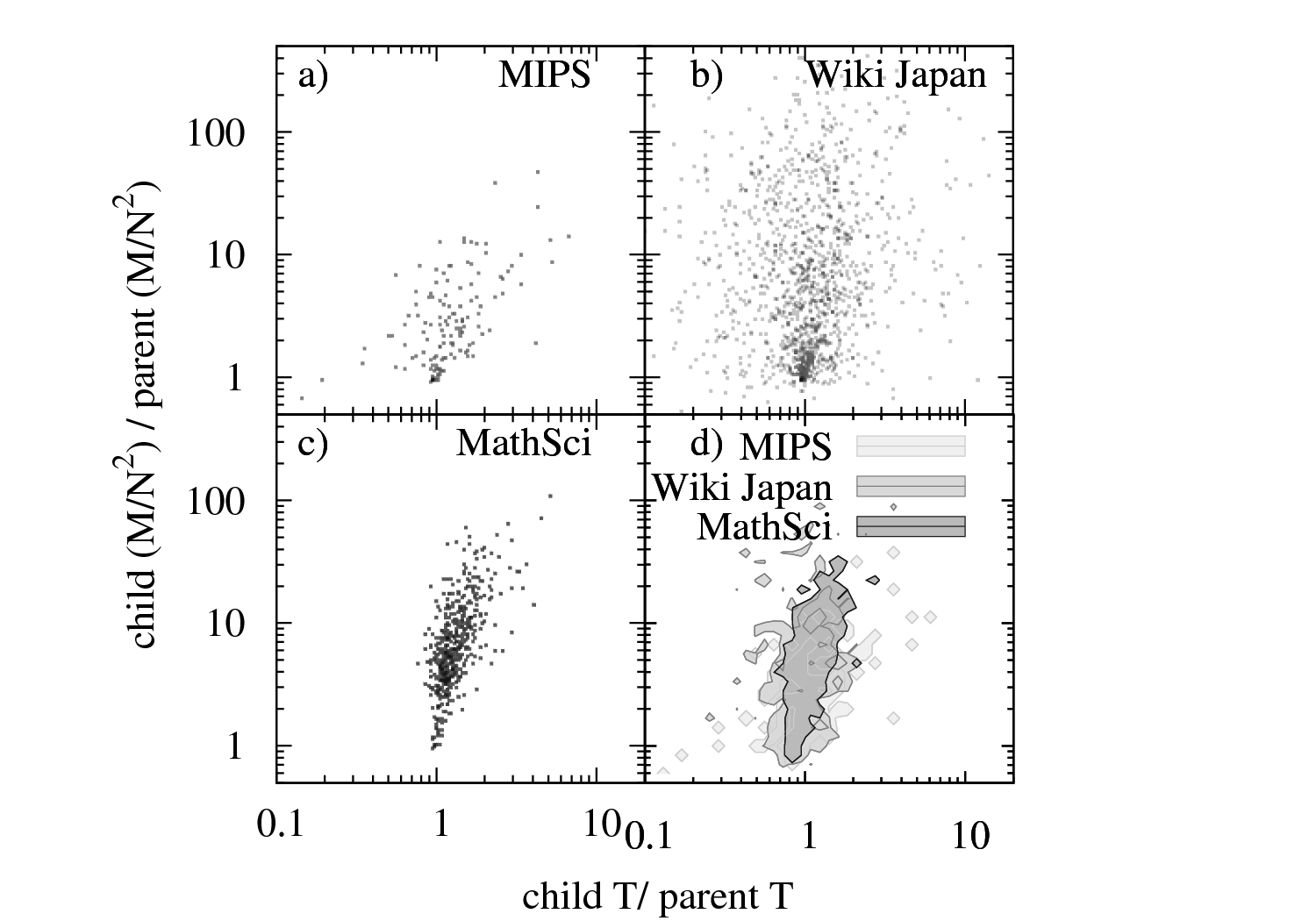}}
\caption{
 The ratio of the direct ancestor-descendant 
link probabilities in function of the ratio of the  
transitivity coefficient for the same pairs in the MIPS (a), the Wiki-Japan (b) 
and the MathScinet networks (c). Similarly to the previous Figs., in panel
 (d) we show a contour plot of the point densities obtained from the 
scatter plots.}
\label{fig:child_ratio}
\end{figure}
Finally, in Fig.\ref{fig:child_ratio}. we plot the ratio of the 
transitivities obtained in the tag-induced sub-graphs corresponding 
to a direct ancestor-descendant pair in function of the ratio of the linking
 probabilities in the same pair of sub-graphs. In case of the MIPS 
(Fig.\ref{fig:child_ratio}a) and the
Wiki-Japan networks (Fig.\ref{fig:child_ratio}b)
 the plots are rather scattered, with a weak increasing
tendency. This means that when moving from the induced sub-graph of
a more generic tag to the induced sub-graph of its direct descendant,
 an increase in the linking probability will more likely  induce
an increase in the transitivity as well, but not necessarily. The points
in case of the MathSciNet (Fig.\ref{fig:child_ratio}c) 
form a much more concentrated cloud than in 
the previous examples, corresponding to a more pronounced positive
correlation between the change in $T$ and the change in $\rho$. This
behavior can be seen in Fig.\ref{fig:child_ratio}d comparing the
 contour plots corresponding to the individual scatter plots.

\section{Summary and conclusions}
\label{sect:concl}
We studied the behavior of the clustering coefficient in tagged networks 
where the tags are organized into an ontology. The investigated systems 
showed universal features in some aspects with interesting differences 
from other perspectives. The average $C$ showed a decaying tendency
in function of the number of tags on the nodes in case of the
MathSciNet and Wiki-Japan networks. This sort of correlation was
absent in case of the MIPS. The tag-induced sub-graphs showed on an 
increasing transitivity coefficient on average 
with decreasing size in all networks. Furthermore,
for the vast majority of the tags the transitivity in the induced sub-graph
was much higher than in an Erd\H{o}s-R{\'e}nyi random graph with the same number
of nodes and links. This difference became really significant towards
the larger sub-graphs, corresponding to more general tags. In other
words, when widening the specificity of tags, 
 the transitivity in the corresponding sub-graphs
decreases at a much slower rate on average then expected based on the
decrease of the linking probability in the sub-graphs. 

We also compared various properties of the tag-induced sub-graphs
to the same properties in the induced sub-graph of the tags direct
ancestor in the DAG. According
to our results for the majority of the ancestor-descendant pairs 
the transitivity and the linking probability is larger in the sub-graph
 of the descendant. The correlation between the transitivity values
 is also stronger than the correlation between the linking probabilities.

\section*{References}

\bibliographystyle{unsrt}
\bibliography{tag_clust}

\end{document}